\documentclass[conference]{IEEEtran}
\ifCLASSINFOpdf
  \usepackage[pdftex]{graphicx}
  \usepackage{multirow}
\else
\fi

\usepackage{url}
\usepackage{cite}
\usepackage{siunitx}
\usepackage{hyperref}
\usepackage[procnames]{listings}
\usepackage{xcolor}
\begin{document}
%

\title{Accelerating FPGA-Based Wi-Fi Transceiver Design and Prototyping by High-Level Synthesis
}


\author{\IEEEauthorblockN{Thijs Havinga,
Xianjun Jiao,
Wei Liu and 
Ingrid Moerman}
\IEEEauthorblockA{\textit{IDLab, Department of Information Technology} \\
\textit{Ghent University - imec}\\
Ghent, Belgium \\
\{firstname.lastname@UGent.be\}}
}


\maketitle

\begin{abstract}
Field-Programmable Gate Array (FPGA)-based Software-Defined Radio (SDR) is well-suited for experimenting with advanced wireless communication systems, as it allows to alter the architecture promptly while obtaining high performance. However, programming the FPGA using a Hardware Description Language (HDL) is a time-consuming task for FPGA developers and difficult for software developers, which limits the potential of SDR. High-Level Synthesis (HLS) tools aid the designers by allowing them to program on a higher layer of abstraction. However, if not carefully designed, it may lead to a degradation in computing performance or significant increase in resource utilization. This work shows that it is feasible to design modern Orthogonal Frequency Division Multiplex (OFDM) baseband processing modules like channel estimation and equalization using HLS without sacrificing performance and to integrate them in an HDL design to form a fully-operational FPGA-based Wi-Fi (IEEE 802.11a/g/n) transceiver. Starting from no HLS experience, a design with minor overhead in terms of latency and resource utilization as compared to the HDL approach was created in less than one month. We show the readability of the sequential logic as coded in HLS, and discuss the lessons learned from the approach taken and the benefits it brings for further design and experimentation. The FPGA design generated by HLS was verified to be bit-true with its MATLAB implementation in simulation. Furthermore, we show its practical performance when deployed on a System-on-Chip (SoC)-based SDR using a professional wireless connectivity tester. 
\end{abstract}

\section{Introduction}
As opposed to Commercial Off-The-Shelf (COTS) wireless chipsets, Software-Defined Radio (SDR) platforms enable researchers to experiment and prototype innovative solutions more freely. However, due to the low-latency and high data rate requirements of current and future communications standards, using a general purpose processor for the baseband signal processing is often not feasible. Even when these high-demanding tasks are offloaded to a hardware accelerator, the interface between the accelerator and the operating host forms a bottleneck \cite{Jiao18}.

A System-on-Chip (SoC), consisting of a Central Processing Unit (CPU) and Field-Programmable Gate Array (FPGA) with high-speed interconnection, is a suitable platform for high-performance systems, while preserving flexibility. Nonetheless, creating Register-Transfer Level (RTL) designs for the FPGA using a Hardware Description Language (HDL) requires explicit specification of each operation on a per clock cycle basis, and its verification is a time-consuming process. High-Level Synthesis (HLS) tools allow the designer to work on a higher level of abstraction in a programming language like (System)C, C++ or OpenCL, which may shorten the design cycle significantly. Naturally, this approach often comes at the cost of additional hardware resources or lower computing performance \cite{Altoyan20}.

In this paper we show that a significant part of the receiver architecture of openwifi \cite{jiao2020openwifi}, an SDR-based IEEE 802.11a/g/n (Wi-Fi) wireless transceiver originally written in Verilog, can be replaced by HLS modules with acceptable overhead. This allows for more flexibility in adapting the existing design, which is required, e.g., to support 802.11ax (Wi-Fi 6), that should be able to handle more configurations in the baseband to support Orthogonal Frequency Divison Multiple Access (OFDMA). HLS also helps to explore and test different algorithms, e.g., for better resistance to multi-path fading in a real-world environment. Converting a high-level design (using e.g. MATLAB) to HLS code requires much less effort than rewriting a large part of the HDL to align the operations on a clock-cycle level. This speeds up the verification process using prototypes in a real-life environment, which is important for wireless research.

Starting from the existing bit-true MATLAB implementation of the HDL design, as designers with no HLS experience, functionally equivalent channel estimation and equalization modules were created in about two weeks. The created parts consist of both data and control paths, such that they operate as standalone modules that can be easily integrated in the existing system. Only little effort in guiding the HLS tool was needed to create a design that meets the latency requirements and could be deployed on a SoC with limited capacity in terms of FPGA resources, to form a fully-operational transceiver. This took again around two weeks. We present the similarities between the MATLAB code that was used as reference, and the HLS code written in C++ using Vitis HLS by AMD Xilinx. This approach of wireless baseband prototyping opens opportunities for wireless system designers that are less familiar with digital hardware design. However, further optimization in terms of hardware utilization requires more effort as will be discussed in later sections. 

The remainder of the paper is organized as follows: first, we present related work on this topic; then, we describe the system architecture, followed by our design flow and lessons learned from this approach; afterwards, we present and discuss the results in terms of resource utilization, computing performance and a validation using a professional wireless connectivity tester. 
\section{Related work}
Several work has been done on using HLS for FPGA-based SDR applications, as shown below. 

WARP Project \cite{warpProject} is an IEEE802.11a/g/n compatible implementation using a dedicated hardware platform. The included FPGA is programmed using Xilinx System Generator, which is generally considered still a rather low-level approach. It blends MATLAB code, Simulink blocks and blackboxes consisting of Verilog code, which makes it difficult to maintain and extend. Transmitter and receiver benchmarks show that the design performs better than required by the standard. The design consumes a relatively large amount of resources, namely 71k Look-Up Tables (LUTs), 80k Flip-Flops (FFs), 216 Block Random Access Memory (BRAM) tiles and 194 Digital Signal Processing blocks (DSPs).

In \cite{Drozdenko18}, multiple hardware/software co-design variants for an IEEE 802.11a transceiver with BPSK modulation and 1/2 coding rate are explored. The authors use MathWorks HDL Coder\textsuperscript{\texttrademark} \cite{MATLABHDLCoder} to generate the programmable logic from a dataflow model using built-in Simulink blocks. They show the execution speed and resource utilization when placing the hardware/software split at different points in the transmitter (Tx) and receiver (Rx) chain. The Rx implementation alone did not fit on the Xilinx ZedBoard \cite{XilinxZedBoard}, but the combined Rx/Tx design fitted on a ZC706 \cite{XilinxZC706}. Evaluation of the designs by bit error rate measurements is left for future work.

In \cite{Bhatnagar13}, the authors explore four different methodologies for FPGA-based SDR design, namely using HDL or HLS, both using either script-based or Graphical User Interface (GUI) tools. An IEEE 802.15.4 (ZigBee) transceiver was implemented using each approach and their features are compared. They conclude that the development time using HLS is reduced considerably. 
For verification, they show the transmitted and received baseband signal, and the spectrum of a single transmitted waveform was compared against a reference signal's spectrum, however also no bit error rate measurement is conducted. Following the conclusions of this work, in \cite{Ouedraogo14} the authors present a Domain Specific Language (DSL) for frame-based modeling of an SDR transceiver. 
A case study based on an IEEE 802.11a transceiver is discussed, but it is not clear how the design is verified and the result is not compared with an HDL implementation.

The work in \cite{Gautier14} is another extension to the aforementioned work. Here the authors explore HLS for implementing parts of an IEEE 802.11g transmitter and 802.15.4 transceiver. More specifically, the proposed FPGA-SDR design flow uses the DSL that leverages a library of signal processing IPs, which a waveform compiler then assembles and converts into an RTL description using HLS tools. Throughput/area tradeoffs when using different iteration intervals and duplication factors for the loops of the correlation bench of the 802.15.4 receiver are visualized using graphs. The 802.15.4 transmitted waveform when the RTL description was synthesized and deployed on an FPGA platform was verified with a Vector Signal Analyzer, which showed the demodulated constellation. For the 802.11g waveform, only its transmitted spectrum is shown.  

The authors of \cite{Akeela2020} present an implementation of the IEEE 802.11a standard using HLS and a hardware/software co-design approach. The different components are compared to full software or hardware designs in terms of processing speed, power consumption and hardware or software cost. While the design fits on the Xilinx ZC706 \cite{XilinxZC706}, there is no evaluation of the overall transmitter or receiver performance.

The work in \cite{cadence} presents the design and verification flow using HLS for the Fast Fourier Transform (FFT) and Viterbi decoding blocks of an IEEE 802.11ah baseband processor. The performance tuning steps taken to arrive at an acceptable design in terms of latency and resource utilization are discussed and it follows that drastic improvements can be achieved when doing this properly.

Other parts of the processing chain of a wireless transceiver that have been designed using HLS are a self-interference canceller \cite{Lahti20}, digital down converter \cite{Sikka21}, encryption engine \cite{Sikka20}, low-density parity check decoder \cite{Andrade17, Mhaske17} and turbo decoder \cite{Cenova19, Conn18, Stirk19}.


In short, most of the works focus on the design methodology to develop a part of the baseband processing using HLS. They provide the resources used and processing time, but mostly did not provide performance analysis for the transceiver itself. Our work advances further as (i) it uses HLS for a core part of an OFDM receiver and integrates it to form a fully-operational transceiver, (ii) it provides thorough performance analysis, and (iii) it is benchmarked against an HDL equivalent design. We believe the steps taken and achievement in this work are valuable for the community when designing future baseband processing modules.
\section{System architecture}
In this work we focus on the receiver baseband architecture of a 20MHz bandwidth Single-Input Single-Output (SISO) Wi-Fi transceiver. As opposed to the transmitter, which has a more straightforward approach, the receiver needs specialized processing blocks to cope with the noise, carrier and sampling frequency offset, and channel impairments of the received signal. Here we describe a typical receiver baseband processing chain for Orthogonal Frequency Division Multiplex (OFDM)-based Wi-Fi systems, partly based on the architecture provided by OpenOFDM \cite{openOFDM}. 

\subsection{Algorithmic specification}
\label{sec:algorithm}
\begin{figure}[h]
    \centering
    \includegraphics{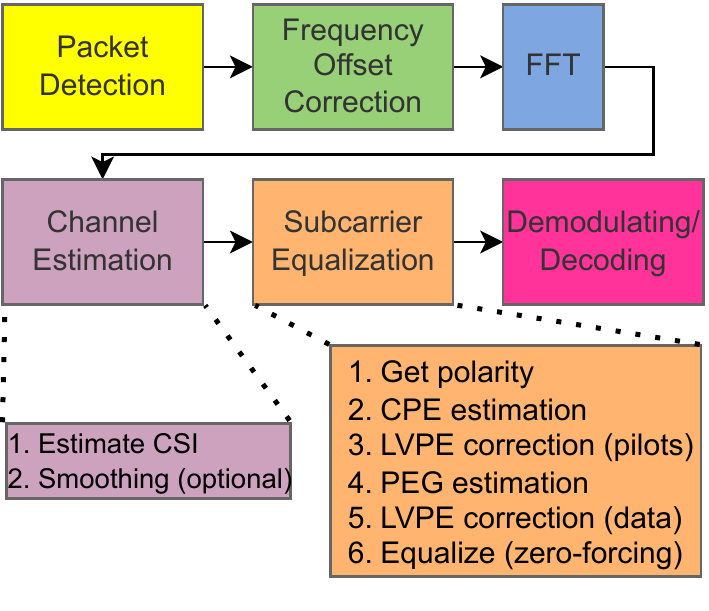}
    \vspace{-0.5cm}
    \caption{Schematic diagram of an OFDM receiver baseband processing chain with the steps of the blocks to be rewritten using HLS.}
    \label{fig:receiver_baseband}
\end{figure}

\begin{figure}
    \centering
    \includegraphics[width=\linewidth]{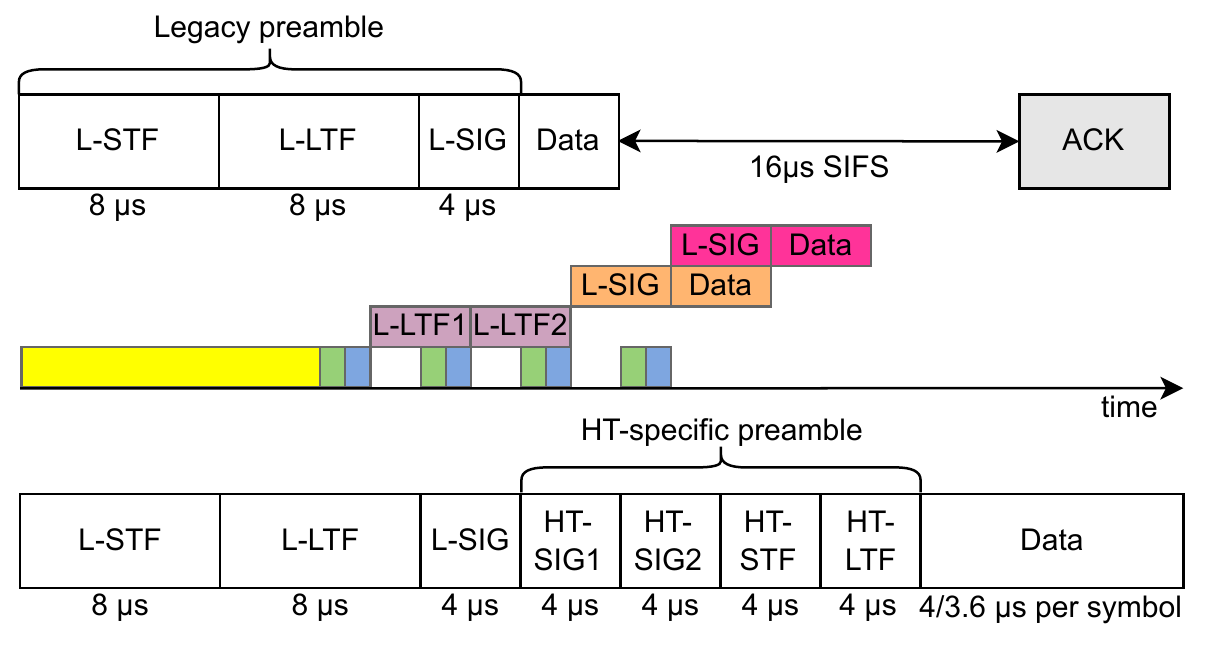}
    \caption{IEEE 802.11 Legacy and HT frame structures together with the processing timeline of Legacy packets. The colored boxes correspond to the blocks in Figure \ref{fig:receiver_baseband}.}
    \label{fig:frame_structure}
    \vspace{-0.2cm}
\end{figure}

A diagram of the receiver's architecture is illustrated in Figure \ref{fig:receiver_baseband}. Figure \ref{fig:frame_structure} shows the frame structure for 802.11a/g (Legacy) and 802.11n (High Throughput, HT) packets. It also illustrates the processing timeline of Legacy packets.

Packet detection is done by calculating the auto correlation metric of the incoming complex I/Q samples with the known sequence of the Legacy Short Training Field (L-STF). The carrier frequency offset is estimated using this sequence and corrected for at each following symbol. After this, the FFT is performed to convert the signal from the time domain to the frequency domain.  

In this work, we specifically targeted the Channel Estimation and Subcarrier Equalization blocks to be implemented using HLS, because these are where most changes are expected for performance improvement or when extending the receiver towards newer Wi-Fi standards. 
These modules are needed to remove the channel impairments and sampling frequency offset from the received signal. Channel estimation is done by comparing the received Long Training Field (LTF) with the reference sequence. In this way, the Channel State Information (CSI) per subcarrier can be estimated: 
\begin{equation}
    H[i] = \mathit{L_R}[i] \times L_T[i], 
\end{equation}
where \textit{H} is the estimated CSI in frequency domain, $L_R$ is the received symbol and $L_T$ is the known sequence (either valued +1 or -1), given at each active subcarrier index \textit{i}. 
For the Legacy part, the calculation is performed by an average of the two received L-LTF symbols, each consisting of 52 active subcarriers. For the HT part, the one-symbol HT-LTF with 56 active subcarriers is used. The indices of all subcarriers range from -31 to 32.

In order to reduce the noise of the CSI, it can be smoothed by a moving average filter spanning multiple subcarriers, but this may lead to distortion under a frequency-selective channel. Therefore, this operation is optional and for HT, a transmitter can recommend doing this by setting a bit in the HT Signal (HT-SIG) field.  

\begin{figure}
    \centering
    \includegraphics{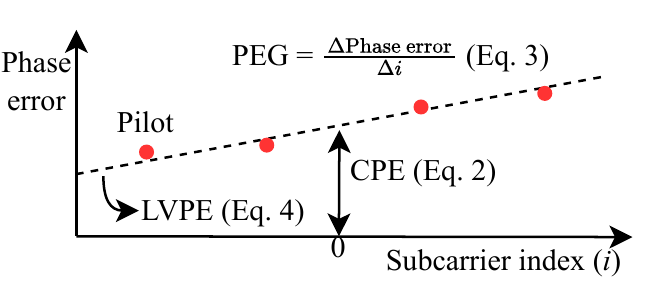}
    \caption{LVPE across the subcarriers given by a CPE and PEG.}
    \label{fig:LVPE}
\end{figure}

For each of the Signal and Data symbols, first the sampling frequency offset is estimated as Linear Varying Phase Error (LVPE) defined by a Common Phase Error (CPE) and Phase Error Gradient (PEG) across the subcarriers, as shown in Figure \ref{fig:LVPE} \cite{CPELVPE}. 
The estimation is done based on the pilot subcarriers, which are modulated using Binary Phase Shift Keying (BPSK), following a pseudo-random known sequence. Therefore, for each symbol, the polarity of the pilots, denoted by $\textit{P}$, has to be traced. As a first step, the CPE, which is the offset of the LVPE for the current symbol, can be calculated as follows: 
\begin{equation}
    \mathit{CPE} = \angle \big(\sum_{i\in I} \overline{X[i]} \times P[i] \times H[i] \big),
\end{equation}
where $\overline{X}$ is the conjugate of the incoming symbol and $I$ are the indices of the pilot subcarriers.
Then the PEG of the LVPE can be derived by linear regression assuming the intercept at subcarrier index 0 is now zero: 
\begin{equation}
    \mathit{PEG} = \frac{ \sum_{i \in I} i \cdot \angle \big( \overline{X[i]} \times P[i] \times H[i] \big)   }{\sum_{i \in I} i^2 }.
\end{equation}
To avoid abnormally large differences in the PEG of adjacent OFDM symbols, the accumulated PEG of the previous symbols can be tracked. Then first the pilots of the current symbol are corrected by the accumulated PEG and only an incremental PEG is estimated and added to the accumulated PEG for the current symbol. 

The LVPE at each data subcarrier is then calculated as: 
\begin{equation}
    \mathit{LVPE}[i] = \mathit{CPE} + i \cdot \mathit{PEG}. 
\end{equation}
Finally, LVPE correction is performed by rotating the subcarriers by the opposite of the resulting phase error.
After this, the symbol corrected for LVPE ($X'$) is equalized. This is done by normalizing the signal with the estimated CSI, which is typically done by zero-forcing:
\begin{equation}
    Y[i] = \frac{X'[i]}{H[i]},
\label{eq:eq_per_sym}
\end{equation}
where \textit{Y} is the resulting symbol that will be fed to the demodulating/decoding block. 

\subsection{Performance criteria}
The aforementioned operations should be accurately organized, especially when they require conditional processing. 
The HLS tool can be of help here, but in order to guide the tool, it is important to know which performance the design should achieve.

\subsubsection{Resource utilization}
In terms of hardware, the targeted SoC with lowest capacity in terms of FPGA resources is the Xilinx Zynq-7000 SoC Z-7020 \cite{XilinxZynq}, which resides for example on the ZedBoard. It has 53,200 LUTs, 106,400 FFs, 140 BRAMs (of 36Kb each) and 220 DSPs. For reference, the utilization of the HDL implementation of the channel estimation and equalizer blocks, as well as the full openwifi design, is given in Table \ref{tab:utilization} in Section \ref{sec:results}.

\subsubsection{Computing performance}
The channel estimation is based on either the L-LTF or HT-LTF. Its result should be available before the data symbol can be equalized. The 64-point FFT block that provides input to the channel estimation and equalization block, processes the incoming I/Q samples (16-bit each), that are sampled at 20MHz, symbol per symbol. Depending on the Guard Interval (GI) used there are 80 (long GI) or 72 (short GI) time-domain samples per symbol, resulting in a symbol duration of 4.0 and 3.6$\mu s$, respectively. This means that the channel estimation result should be available within 3.6$\mu s$, such that it can be used for the symbol coming afterwards.

The main performance criterion for the equalizer is a minimum throughput as induced by the throughput of the FFT block. After removing the GI, there are 64 frequency domain samples per symbol, generated in one burst at one sample per clock cycle, that need to be processed. Therefore, the equalizer should be able to handle 64 samples per symbol duration, resulting in 17.8 samples per $\mu s$ when using a short GI. Next to this, the overall decoding latency should be kept low such that an acknowledgement (ACK) to a received frame can be sent after the Short Interframe Space (SIFS) of 16$\mu s$ \cite{IEEE20}.

\section{Approach}
\subsection{Design flow}
\label{sec:design}
As starting point we use a bit-true model of the HDL design which is implemented in MATLAB. However, there remains a significant amount of work to make it compatible with MATLAB HDL Coder's input requirement \cite{MATLABHDLCoder}, which makes this approach not attractive. Since our target platforms are Xilinx boards, we chose to use C++ within Vitis HLS by AMD Xilinx \cite{VitisHLS}, but the approach taken should be generic for any HLS tool. Converting the algorithms from MATLAB to C++ requires minor effort (around three days in our case). In HLS we use Arbitrary Precision (AP) data types that correspond to the HDL implementation, which are emulated by truncating and rounding operations in MATLAB. Input and output files to be used with testbenches are created with MATLAB to verify the correctness of the algorithms using C simulation. 

After inserting pragmas to optimize performance and resource utilization, C/RTL co-simulations are performed to confirm that the design meets the requirements. After integrating the HLS generated parts with the rest of the HDL-based baseband processing, the complete RTL receiver design is verified in simulation by generating packets with random configurations in MATLAB and reading them in the HDL-based testbench. The HLS generated RTL design is confirmed to be bit-true with the MATLAB implementation. It proved that having such a reference is highly desired not only for the initial design phase, but also to discover discrepancies before deployment.

The open-source code is available at \cite{hlscode}. An application node of how to use the HLS code with openwifi is given at \cite{applicationnote}. As an example, we present the HLS code of the main loop of the top-level equalizer function, as shown in Listing \ref{lst:equalizer}.
\begin{lstlisting}[caption={HLS code for the equalizer loop.},label={lst:equalizer}]
cplx16 symbol_iq[64]; int16 cpe;cplx16 lts_pilot[4];
uint1 current_polarity[4]; int18 sym_phase[64];
int24 Sxy; uint7 length, pol_nr; uint6* arr;
for (uint12 i=0; i<nof_ofdm_sym; i++) {
#pragma HLS ALLOCATION function instances=lvpe_correction limit=1
    for (uint7 j=0; j<64; j++) {
#pragma HLS PIPELINE
        symbol_iq[j] = sample_in.read();
    }
    pol_nr = get_polarity(current_polarity, ht, pol_nr);
    cpe = cpe_estimate(symbol_iq, lts_pilot, current_polarity);
    lvpe_correction(sym_phase, cpe, acc_PEG, 0, pilot_loc, 4); // pilots only
    Sxy = peg_estimate(symbol_iq, sym_phase, lts_pilot, current_polarity);
    acc_PEG = lvpe_correction(sym_phase, cpe, acc_PEG, Sxy, arr, length); // data 
    equalizer_per_sym(symbol_iq, sym_phase, sample_out, new_lts, arr, length);
}
\end{lstlisting}
Like in the MATLAB specification, the subfunctions (highlighted in red) inside the equalizer are called sequentially. This significantly enhances the comprehensibility of the code. Also, no specific timing related logic resides in the code as this task is offloaded to the HLS scheduler. Therefore, adding functionality or modifying the algorithms does not require manual adjustments regarding timing, as would be the case using an HDL approach. 

\subsection{Optimization strategies}
Next we list some of the lessons learned from the optimization process. The HLS \texttt{dataflow} pragma (task-level parallelism) was  not applied in the top-level function, because there is data-dependency between the functions. All loops in the subfunctions itself are pipelined with the default iteration interval of 1 clock cycle, by directing the HLS tool to do so in order to make sure the design meets the latency requirements. Hardware duplication of the \textit{lvpe\_correction} instance was avoided by limiting its allocation to one. The phase calculation and rotation of I/Q samples are realized by lookup tables for the sine, cosine and arc-tangent functions with limited integer resolution, which is in accordance to the original openwifi HDL implementation. Using the HLS Math Library for these trigonometric functions would lead to high resource utilization with unnecessary high resolution.


The design consists of several division operations, which require a lot of hardware, especially when low latency is required. For example, the division from Equation \ref{eq:eq_per_sym} should be done for both the real and imaginary part of an FFT output. When using the HLS \texttt{allocation} pragma to limit this \texttt{sdiv} operation to one, the design could already not meet the latency requirements. This is because it increased the iteration interval of the loop through all subcarriers considerably, as it could not parallelize the two operations anymore. This optimization is therefore discarded.

To ease the HLS scheduler's work, conditions inside the subfunctions should be avoided as much as possible. Therefore, although the incoming subcarriers arrive in order, the best performance was obtained when writing them all to an array (which gets synthesized as a BRAM), because out of the 64 incoming subcarriers only the active ones need to be processed. A static array containing the indices of the active subcarriers, which will be synthesized as Read-Only Memory (ROM), is then used to read from the array. This removes the need for conditional statements in the loops, while only the number of loop iterations (as given by the parameter \textit{length}) varies. The latter is well handled by the HLS scheduler.  

Moreover, since the operations of multiple subfunctions depend on whether the Legacy or HT part of a packet is handled, the best performance was achieved when evaluating this condition before the main \textit{equalize} loop. This requires to restart the equalizer module when switching from Legacy to HT, which has to be done when integrating the modules in the existing HDL design. Integration requires minor effort, as Vitis HLS automatically generates suitable interfaces for the block-level handshakes it generates and function arguments defined in C++. This allows us to easily restart the HLS modules and provide the necessary inputs in the top-level HDL state machine, which can be found in \textit{verilog/dot11.v} in \cite{hlscode}.   
\section{Results and discussion}
\label{sec:results}
\begin{table*}[]
\centering
\caption{Hardware utilization when using HDL or HLS methodology.}
\label{tab:utilization}
\resizebox{\textwidth}{!}{%
\begin{tabular}{|l|l|ll|ll|ll|ll|}
\hline
Design &
  \multicolumn{1}{c|}{Methodology} &
  \multicolumn{2}{l|}{LUT} &
  \multicolumn{2}{l|}{FF} &
  \multicolumn{2}{l|}{BRAM} &
  \multicolumn{2}{l|}{DSP} \\ \hline
\multirow{3}{*}{\begin{tabular}[c]{@{}l@{}}Channel Estimation\\ and Equalizer\end{tabular}} &
  HDL &
  \multicolumn{1}{l|}{5,417} &
  100\% &
  \multicolumn{1}{l|}{10,073} &
  100\% &
  \multicolumn{1}{l|}{1} &
  100\% &
  \multicolumn{1}{l|}{19} &
  100\% \\ \cline{2-10} 
 &
  HLS &
  \multicolumn{1}{l|}{6,833} &
  \textcolor{red}{126.1\%} &
  \multicolumn{1}{l|}{7,208} &
  \color{black!40!green}71.6\%\color{black} &
  \multicolumn{1}{l|}{5} &
  \textcolor{red}{500\%} &
  \multicolumn{1}{l|}{34} &
  \textcolor{red}{178.9\%} \\ \cline{2-10} 
 &
  HLS\textsuperscript{1} &
  \multicolumn{1}{l|}{5201} &
  \color{black!40!green}96.0\%\color{black} &
  \multicolumn{1}{l|}{-} &
  - &
  \multicolumn{1}{l|}{3.5} &
  \textcolor{red}{350\%} &
  \multicolumn{1}{l|}{25} &
  \textcolor{red}{131.6\%} \\ \hline
\multirow{2}{*}{openwifi} &
  HDL &
  \multicolumn{1}{l|}{34,630 (64.6\%)} &
  100\% &
  \multicolumn{1}{l|}{49,168 (46.2\%)} &
  100\% &
  \multicolumn{1}{l|}{111 (79.3\%)} &
  100\% &
  \multicolumn{1}{l|}{111 (50.5\%)} &
  100\% \\ \cline{2-10} 
 &
  HLS &
  \multicolumn{1}{l|}{36,867 (69.3\%)} &
  \textcolor{red}{106.5\%} &
  \multicolumn{1}{l|}{48,815 (45.9\%)} &
  \color{black!40!green}99.3\%\color{black} &
  \multicolumn{1}{l|}{115 (82.1\%)} &
  \textcolor{red}{103.6\%} &
  \multicolumn{1}{l|}{126 (57.3\%)} &
  \textcolor{red}{113.5\%} \\ \hline
\end{tabular}%
}
\vspace{0.1cm}

\textsuperscript{1}Derived from Vivado implementation report when using dedicated IP for the dividers and other resource optimizations (see Section \ref{sec:util}).
\end{table*}

\begin{table}[!h]
\centering
\caption{Latency of the Equalizer and openwifi using HDL or HLS.}
\label{tab:latency}
\resizebox{0.4\textwidth}{!}{%
\begin{tabular}{|l|l|l|l|}
\hline
Design & \multicolumn{1}{c|}{Methodology} & \begin{tabular}[c]{@{}l@{}}Latency \\ Legacy ($\mu s$)\end{tabular} & \begin{tabular}[c]{@{}l@{}}Latency \\ HT ($\mu s$)\end{tabular} \\ \hline
\multirow{2}{*}{Equalizer} & HDL & 2.52  & 2.52 \\ \cline{2-4} 
                           & HLS & 2.82  & 2.90 \\ \hline
\multirow{2}{*}{openwifi}  & HDL & 8.79 & 8.89 \\ \cline{2-4} 
                           & HLS & 9.09 & 9.27 \\ \hline
\end{tabular}%
}
\vspace{-0.2cm}
\end{table}


\subsection{Hardware utilization}
\label{sec:util}
We evaluate the designs created using both HDL and HLS in terms of their hardware utilization on a Xilinx ZedBoard \cite{XilinxZedBoard}, at a clock frequency of 100MHz. We compare the amount of LUTs, FFs, BRAMs and DSPs after performing implementation in Vivado. We show the number of required parts for each resource type in Table \ref{tab:utilization}. The resource utilization reported by Vivado after implementation is given in percentage relative to the HDL approach, highlighted by either red or green digits, depending on whether the usage is increased. For the complete openwifi design, we also present the utilized percentage of the total available resources on the ZedBoard.

Among the four types of FPGA resources, except for the number of FFs, the utilization of the rest resources has increased by the HLS modules. The extra LUTs consumption is mostly due to the divider instances that are used in pipelined loops in order to meet the latency requirements. Three divider instances take up 4,636 out of the 6,833 LUTs. 
This could not be shrunk by e.g. increasing the iteration interval. As a comparison, these dividers in the HDL design are implemented with a specific IP core, that provides comparable performance, but takes up only 3,004 LUTs. 
Thus, if it would be possible to use these dedicated IPs, the HLS design ends up with 5,201 LUTs, which is less than the HDL implementation's consumption. 

The increase in BRAMs usage of the HLS generated design is due to the fact that arrays are used to store raw and processed I/Q samples, as explained in Section \ref{sec:design}. In HDL these are mostly handled in a streaming manner, leading to 1 additional BRAM for channel estimation and 1.5 (3 16Kb-tiles) for equalization. On the other hand, Vitis HLS automatically uses BRAMs for the implementation of ROMs, but these can be converted to LUTs by using the HLS \texttt{bind\_storage} pragma, which would save 1.5 BRAMs according to HLS implementation. 

Restricting the number of complex multiplication instances across multiple subfunctions is possible using the HLS \texttt{allocation} pragma in combination with function pointers. This can further decrease the number of DSPs by 4, at the cost of some minor increase in LUTs and FFs according to HLS implementation. Removing pipelining or increasing its iteration interval in some of the smaller loops (e.g. through the pilot subcarriers) is possible, which results in slightly lower LUT, FF and DSP utilization at the cost of some additional latency. The amount of DSPs can be reduced by 3 in this way, but when eliminating another 6 DSPs by removing pipelining in larger loops, the throughput requirements could not be met. The remaining additional 2 DSP modules used by HLS are induced to perform a constant division, whereas a divider module is used in HDL, which uses LUTs and FFs. 

The impact on the total utilization of openwifi is limited. It is slightly different than the difference between the channel estimation and equalization modules due to to the logic to integrate them. Furthermore, the phase calculation instance as used in the equalizer is shared with another module in the HDL variant, which saves LUTs and FFs.  

\subsection{Throughput and latency}
The latency of the channel estimation is measured in simulation by the time between the last I/Q sample of a symbol is received at the input and produced at the output. We measure the worst-case latency (including smoothing) to be 1.62$\mu s$ for the Legacy part and 1.66$\mu s$ for the HT part (due to more active subcarriers), which is well below the required 3.6$\mu s$. 

Next, we evaluate the equalizer and full openwifi design in terms of latency for a Legacy and HT packet (with MCS 0). For the openwifi design, the output is considered ready when the result of the Frame Check Sequence (FCS) is available. This is shown in Table \ref{tab:latency}. 

It can be seen that using HLS, the latency of the equalizer is increased by around 0.30$\mu s$ per OFDM symbol for Legacy packets and 0.38$\mu s$ for HT packets. The equalizer cannot accept new samples after the full symbol is processed, meaning that the burst of 64 samples at 100MHz from the FFT needs to be added to this latency, leading to a worst-case throughput of 18.1 samples per $\mu s$, which is higher than the required 17.8 samples per $\mu s$.


As expected, the increased latency of the equalizer leads to the same increase for the overall openwifi design. This limited increase in latency is acceptable since the design can still meet the requirement of sending an ACK in time. Note that the total latency depends on the number of bytes in the last OFDM symbol due to the decoding process afterwards, which is outside the scope of this paper. 

\subsection{Practical validation}
The FPGA design with the HLS generated channel estimator and equalizer is loaded on the ZedBoard and correctly interacts with the Linux mac80211 subsystem running on the on-chip CPU. It proves to be functional as it can find the available Wi-Fi networks in the neighborhood, and a smartphone can connect when the board operates as an Access Point (AP). 

Next, we test the receiver performance of both the HDL and HLS-enhanced design using a professional wireless connectivity tester -- the R\&S\textsuperscript{\tiny\textregistered}CMW270 \cite{CMW}. The tester is set as an AP in 802.11g(OFDM)/n 20MHz Signaling mode and the board acts as a Station controlled by \texttt{wpa\_supplicant}. The board is connected to the tester using a coaxial cable with known loss of 1.3dB. The receiver sensitivity tests are performed by letting the tester generate 10,000 HT packets, each with 512 bytes as payload. Upon successful decoding, the board sends back an acknowledgement. The PER is calculated as the ratio of the unacknowledged packets over the total transmitted packets. The receiver sensitivity is defined as the received power at which no more than 10\% PER is achieved. This point is found by changing the transmit power of the tester, from which we derive the received power by compensating for the cable loss. The results for the different Modulation and Coding Schemes (MCS) are shown in Table \ref{tab:receiver_sensitivity}. For reference, the achieved PER is given, as well as the receiver sensitivity of a COTS chip \cite{NXP} in 802.11n 20MHz mode. 

\begin{table}[]
\centering
\caption{Receiver sensitivity performance of the HDL and HLS design.}
\label{tab:receiver_sensitivity}
\resizebox{\linewidth}{!}{%
\begin{tabular}{l|ll|ll|l}
\cline{2-6}
                        & \multicolumn{2}{l|}{HDL}           & \multicolumn{2}{l|}{HLS}     &  \multicolumn{1}{l|}{COTS \cite{NXP}}    \\ \hline
\multicolumn{1}{|l|}{MCS} &
  \multicolumn{1}{l|}{\begin{tabular}[c]{@{}l@{}}Received\\ power (dBm)\end{tabular}} &
  PER (\%) &
  \multicolumn{1}{l|}{\begin{tabular}[c]{@{}l@{}}Received\\ power (dBm)\end{tabular}} &
  PER (\%) & \multicolumn{1}{l|}{\begin{tabular}[c]{@{}l@{}}Receiver\\ sensitivity\end{tabular}}\\ \hline
\multicolumn{1}{|l|}{0} & \multicolumn{1}{l|}{-92.80}  & 9.26 & \multicolumn{1}{l|}{-92.50} & 9.51 & \multicolumn{1}{l|}{-89 dBm} \\ \hline
\multicolumn{1}{|l|}{1} & \multicolumn{1}{l|}{-90.70}  & 9.73 & \multicolumn{1}{l|}{-90.40} & 9.36 & \multicolumn{1}{l|}{-88 dBm} \\ \hline
\multicolumn{1}{|l|}{2} & \multicolumn{1}{l|}{-88.65} & 9.72 & \multicolumn{1}{l|}{-88.55} & 9.75 & \multicolumn{1}{l|}{-85 dBm} \\ \hline
\multicolumn{1}{|l|}{3} & \multicolumn{1}{l|}{-86.05} & 9.71 & \multicolumn{1}{l|}{-85.80} & 9.83 & \multicolumn{1}{l|}{-83 dBm} \\ \hline
\multicolumn{1}{|l|}{4} & \multicolumn{1}{l|}{-82.85} & 9.51 & \multicolumn{1}{l|}{-82.75} & 9.66 & \multicolumn{1}{l|}{-80 dBm} \\ \hline
\multicolumn{1}{|l|}{5} & \multicolumn{1}{l|}{-77.70} & 9.21 & \multicolumn{1}{l|}{-77.60} & 9.76 & \multicolumn{1}{l|}{-76 dBm} \\ \hline
\multicolumn{1}{|l|}{6} & \multicolumn{1}{l|}{-76.05} & 9.50 & \multicolumn{1}{l|}{-76.05} & 9.99 & \multicolumn{1}{l|}{-74 dBm} \\ \hline
\multicolumn{1}{|l|}{7} & \multicolumn{1}{l|}{-73.30} & 9.94 & \multicolumn{1}{l|}{-73.30} & 9.83 & \multicolumn{1}{l|}{-73 dBm} \\ \hline
\end{tabular}%
}
\vspace{-0.2cm}
\end{table}

It can be seen that the HDL-based and HLS-enhanced designs show very comparable receiver sensitivity performance, which is the anticipated result since they have similar functional architecture. 
\section{Conclusions and future work}
We explore the feasibility of using HLS to create baseband processing modules for FPGA-based SDR applications in order to accelerate the design flow. Starting from a MATLAB implementation that is bit-true with the original HDL design, a functionally equivalent C++ implementation of the channel estimation and equalizer algorithms is created using the Vitis HLS tool in about two weeks. After refactoring and inserting pragmas to guide the tool, RTL designs are generated that meet the throughput and latency requirements set upfront. These are then integrated with a full HDL-based Wi-Fi receiver baseband design, which is verified using RTL simulations. Implementation results show a minor increase in overall resource utilization, still allowing to accommodate the design on the originally supported board with the least resources. Finally, the HLS-enhanced transceiver design is deployed on a Xilinx ZedBoard and proves to have similar performance in terms of receiver sensitivity when compared to its HDL counterpart using a wireless connectivity tester. Thus, we have shown that using HLS in the design flow for FPGA-based SDR can significantly reduce the development time and speed up prototyping. However, understanding what hardware is generated for different software architectures is still an important aspect for creating an efficient design, as shown by the optimization steps that we performed. 

For future work we will investigate using dedicated HDL IPs in HLS, e.g., by wrapping them as RTL blackbox, and to let different HDL modules use the same integrated HLS subfunction in order to reduce hardware utilization. Next we plan to further enhance the design for better resistance to multipath fading and experiment with it, which is now more attainable with the infrastructure we created by converting a significant part to HLS. It will also ease the implementation of more advanced standards like Wi-Fi 6 and 7. 




\section*{Acknowledgment}
This research was partially funded by the Flemish FWO SBO \#S003921N VERI-END.com project.



\bibliographystyle{IEEEtran}
\bibliography{IEEEabrv,references.bib}

\begin{thebibliography}{10}
\providecommand{\url}[1]{#1}
\csname url@samestyle\endcsname
\providecommand{\newblock}{\relax}
\providecommand{\bibinfo}[2]{#2}
\providecommand{\BIBentrySTDinterwordspacing}{\spaceskip=0pt\relax}
\providecommand{\BIBentryALTinterwordstretchfactor}{4}
\providecommand{\BIBentryALTinterwordspacing}{\spaceskip=\fontdimen2\font plus
\BIBentryALTinterwordstretchfactor\fontdimen3\font minus
  \fontdimen4\font\relax}
\providecommand{\BIBforeignlanguage}[2]{{%
\expandafter\ifx\csname l@#1\endcsname\relax
\typeout{** WARNING: IEEEtran.bst: No hyphenation pattern has been}%
\typeout{** loaded for the language `#1'. Using the pattern for}%
\typeout{** the default language instead.}%
\else
\language=\csname l@#1\endcsname
\fi
#2}}
\providecommand{\BIBdecl}{\relax}
\BIBdecl

\bibitem{Jiao18}
X.~Jiao, I.~Moerman, W.~Liu, and F.~A.~P. de~Figueiredo, ``{Radio Hardware
  Virtualization for Coping with Dynamic Heterogeneous Wireless
  Environments},'' in \emph{Cognitive Radio Oriented Wireless Networks}.\hskip
  1em plus 0.5em minus 0.4em\relax Cham: Springer International Publishing,
  2018, pp. 287--297.

\bibitem{Altoyan20}
W.~Altoyan and J.~J. Alonso, ``{Investigating Performance Losses in High-Level
  Synthesis for Stencil Computations},'' in \emph{2020 IEEE 28th Annual
  International Symposium on Field-Programmable Custom Computing Machines
  (FCCM)}, 2020, pp. 195--203.

\bibitem{jiao2020openwifi}
X.~Jiao, W.~Liu, M.~Mehari, M.~Aslam, and I.~Moerman, ``{openwifi: a free and
  open-source IEEE802.11 SDR implementation on SoC},'' in \emph{2020 IEEE 91st
  Vehicular Technology Conference (VTC2020-Spring)}.\hskip 1em plus 0.5em minus
  0.4em\relax IEEE, 2020, pp. 1--2.

\bibitem{warpProject}
``{WARP Project},'' \url{http://warpproject.org}, accessed: January 9th, 2023.

\bibitem{Drozdenko18}
B.~Drozdenko, M.~Zimmermann, T.~Dao, K.~Chowdhury, and M.~Leeser,
  ``{Hardware-Software Codesign of Wireless Transceivers on Zynq Heterogeneous
  Systems},'' \emph{IEEE Transactions on Emerging Topics in Computing}, vol.~6,
  no.~4, pp. 566--578, 2018.

\bibitem{MATLABHDLCoder}
``{MATLAB HDL Coder},'' \url{https://nl.mathworks.com/products/hdl-coder.html},
  accessed: December 2nd, 2022.

\bibitem{XilinxZedBoard}
Xilinx, ``{ZedBoard},''
  \url{https://www.xilinx.com/products/boards-and-kits/1-8dyf-11.html},
  accessed: January 9th, 2023.

\bibitem{XilinxZC706}
------, ``{Xilinx Zynq-7000 SoC ZC706 Evaluation Kit},''
  \url{https://www.xilinx.com/products/boards-and-kits/ek-z7-zc706-g.html},
  accessed: January 9th, 2023.

\bibitem{Bhatnagar13}
V.~Bhatnagar, G.~S. Ouedraogo, M.~Gautier, A.~Carer, and O.~Sentieys, ``{An
  FPGA Software Defined Radio Platform with a High-Level Synthesis Design
  Flow},'' in \emph{2013 IEEE 77th Vehicular Technology Conference (VTC
  Spring)}, 2013, pp. 1--5.

\bibitem{Ouedraogo14}
G.~S. Ouedraogo, M.~Gautier, and O.~Sentieys, ``{Frame-based modeling for
  automatic synthesis of FPGA-Software Defined Radio},'' in \emph{{2014 9th
  International Conference on Cognitive Radio Oriented Wireless Networks and
  Communications (CROWNCOM)}}, 2014, pp. 341--346.

\bibitem{Gautier14}
M.~Gautier, G.~S. Ouedraogo, and O.~Sentieys, ``{Design Space Exploration in an
  FPGA-Based Software Defined Radio},'' in \emph{2014 17th Euromicro Conference
  on Digital System Design}, 2014, pp. 22--27.

\bibitem{Akeela2020}
R.~Akeela and Y.~Elziq, ``{Efficient co-design partitioning of WLANs on
  SoC-based SDRs},'' \emph{Microsystem Technologies}, vol.~26, 04 2020.

\bibitem{cadence}
F.~Mighani, M.~Sharp, M.~McNamara, and D.~Pursley, ``{Using High-Level
  Synthesis to Design and Verify 802.11ah Baseband IP},''
  \url{https://www.cadence.com/content/dam/cadence-www/global/en_US/documents/tools/digital-design-signoff/adapt-ip-hls-wp.pdf},
  Cadence, Tech. Rep., November 2016.

\bibitem{Lahti20}
S.~Lahti, P.~Pascual~Campo, V.~Lampu, L.~Anttila, M.~Valkama, and
  T.~Hamalainen, ``{Implementation of a Nonlinear Self-Interference Canceller
  using High-Level Synthesis},'' 10 2020, pp. 1--5.

\bibitem{Sikka21}
\BIBentryALTinterwordspacing
P.~Sikka, A.~R. Asati, and C.~Shekhar, ``{Power- and Area-Optimized High-Level
  Synthesis Implementation of a Digital Down Converter for Software-Defined
  Radio Applications},'' \emph{Circuits Syst. Signal Process.}, vol.~40, no.~6,
  p. 2883–2894, jun 2021. [Online]. Available:
  \url{https://doi.org/10.1007/s00034-020-01601-9}
\BIBentrySTDinterwordspacing

\bibitem{Sikka20}
\BIBentryALTinterwordspacing
------, ``{Speed optimal FPGA implementation of the encryption algorithms for
  telecom applications},'' \emph{Microprocessors and Microsystems}, vol.~79, p.
  103324, 2020. [Online]. Available:
  \url{https://www.sciencedirect.com/science/article/pii/S014193312030483X}
\BIBentrySTDinterwordspacing

\bibitem{Andrade17}
J.~Andrade, N.~George, K.~Karras, D.~Novo, F.~Pratas, L.~Sousa, P.~Ienne,
  G.~Falcao, and V.~Silva, ``{Design Space Exploration of LDPC Decoders Using
  High-Level Synthesis},'' \emph{IEEE Access}, vol.~5, pp. 14\,600--14\,615,
  2017.

\bibitem{Mhaske17}
S.~Mhaske, H.~Kee, T.~Ly, A.~Aziz, and P.~Spasojevic, ``{FPGA-Based Channel
  Coding Architectures for 5G Wireless Using High-Level Synthesis},''
  \emph{International Journal of Reconfigurable Computing}, vol. 2017, pp.
  1--23, 06 2017.

\bibitem{Cenova19}
T.~Cenova, ``{Exploring HLS Coding Techniques to Achieve Desired Turbo Decoder
  Architectures},'' Ph.D. dissertation, Rochester Institute of Technology,
  2019, accessed from: \url{https://scholarworks.rit.edu/theses/10256}.

\bibitem{Conn18}
B.~E. Conn, ``{Exploring High Level Synthesis to Improve the Design of Turbo
  Code Error Correction in a Software Defined Radio Context},'' Ph.D.
  dissertation, Rochester Institute of Technology, 2018, accessed from:
  \url{https://scholarworks.rit.edu/theses/9839}.

\bibitem{Stirk19}
W.~Stirk and J.~Goeders, ``{Implementation and Design Space Exploration of a
  Turbo Decoder in High-Level Synthesis},'' in \emph{2019 International
  Conference on ReConFigurable Computing and FPGAs (ReConFig)}, 2019, pp. 1--5.

\bibitem{openOFDM}
J.~Shi, ``{OpenOFDM: Synthesizable, Modular Verilog Implementation of 802.11
  OFDM Decoder},'' \url{https://openofdm.readthedocs.io/en/latest/}, 2017,
  accessed: December 5th, 2022.

\bibitem{CPELVPE}
MathWorks, ``{Joint Sampling Clock and Carrier Frequency Offset Tracking},''
  \url{https://nl.mathworks.com/help/wlan/ug/joint-sampling-rate-and-carrier-frequency-offset-tracking.html},
  2022, accessed: December 16th, 2022.

\bibitem{XilinxZynq}
``{Zynq-7000 SoC Data Sheet: Overview},''
  \url{https://docs.xilinx.com/v/u/en-US/ds190-Zynq-7000-Overview}, accessed:
  December 5th, 2022.

\bibitem{IEEE20}
``{IEEE Standard for Information technology - Part 11: Wireless LAN Medium
  Access Control (MAC) and Physical Layer (PHY) Specifications},'' \emph{IEEE
  Std 802.11-2020 (Revision of IEEE Std 802.11-2016)}, pp. 1--4379, 2021.

\bibitem{VitisHLS}
``{Xilinx Vitis High-Level Synthesis},''
  \url{https://docs.xilinx.com/r/en-US/ug1399-vitis-hls}, accessed: December
  2nd, 2022.

\bibitem{hlscode}
\url{https://github.com/open-sdr/openofdm/tree/dot11zynq_hls}.

\bibitem{applicationnote}
T.~Havinga, ``{Application Note for using openwifi with HLS.}''
  \url{https://github.com/open-sdr/openwifi/blob/master/doc/app_notes/hls.md}.

\bibitem{CMW}
\emph{{R\&S CMW Wideband Radio Communication Tester}}, Rohde \& Schwarz, 2019,
  {Version 05.00}.

\bibitem{NXP}
\emph{{2.4/5 GHz Dual-band 1x1 Wi-Fi 5 (802.11ac) and Bluetooth 5.2 Solution}},
  NXP, 2022, {Rev. 6}.

\end{thebibliography}

\end{document}